# Exploring the limits of ultracold atoms in space

RJ Thompson[1], D.C. Aveline[1] Sheng-Wey Chiow[1], ER Elliott[1], JR Kellogg[1], JM Kohel[1], MS Sbroscia[1], C. Schneider[1], JR Williams[1], N. Lundblad[2], CA Sackett[3], D. Stamper-Kurn[4,5,6], and L. Woerner[7]

[1] Jet Propulsion Laboratory, California Institute of Technology, Pasadena, CA 91109 USA
[2] Department of Physics and Astronomy, Bates College, Lewiston, ME, 04240, USA
[3] Department of Physics, University of Virginia, Charlottesville, VA, 22904, USA
[4] Department of Physics, University of California, Berkeley, California 94720, USA
[5] Challenge Institute for Quantum Computation, University of California, Berkeley, California 94720, USA
[6] Materials Sciences Division, Lawrence Berkeley National Laboratory, Berkeley, California 94720, USA
[7] German Aerospace Center for Space Systems, DLR-RY, Linzerstrasse 1, D-28359 Bremen, Germany.

E-mail: Robert.J.Thompson@JPL.NASA.GOV

**Abstract**

Existing space-based cold atom experiments have demonstrated the utility of microgravity for improvements in observation times and for minimizing the expansion energy and rate of a freely evolving coherent matter wave. In this paper we explore the potential for space-based experiments to extend the limits of ultracold atoms utilizing not just microgravity, but also other aspects of the space environment such as exceptionally good vacuums and extremely cold temperatures. The tantalizing possibility that such experiments may one day be able to probe physics of quantum objects with masses approaching the Plank mass is discussed.

## 1. Introduction

The successful production and study of atomic Bose-Einstein condensation aboard the MAIUS sounding rocket mission [1], and the continuously operating Cold Atom Laboratory (CAL) user facility aboard the International Space Station (ISS) [2], demonstrate that ultracold atomic physics research can be conducted in freely falling experimental setups. These experiments exploit absence of differential gravitational acceleration between ultracold atoms evolving freely within a vacuum chamber and the vacuum chamber itself. That is, in the absence of any deliberately applied forces, a quantum gas remains inertially confined within the observation volume of the experimental setup. Experiments performed within these setups have exploited this microgravity feature, permitting, for example, long observation times for freely expanding Bose-Einstein condensed gases, enhanced by atom-optical manipulations that minimize the expansion energy of these gases to the pico-Kelvin energy range [3,4], Other experiments have exploited microgravity to impose novel trapping geometries for ultracold atoms -- spherical-shell (bubble) potentials produced by radio-frequency dressing of magnetic trapping potentials -- that would otherwise be strongly distorted by gravitational sag [5].

A comprehensive research agenda targeted at ultracold atomic and molecular gases in microgravity has been envisioned, and this vision is guiding the development of the CAL and its potential upgrades, and of the Bose-Einstein Condensate and Cold Atom Lab (BECCAL) joint mission of NASA, and the German space agency (DLR). [6] As discussed elsewhere [7], the background-potential-free environment within a freely-falling ultracold-atom experimental apparatus opens up several compelling research directions. These include the development of atom interferometers with enhanced interrogation times and exploiting the ability to inertially confine matter waves near physical objects; studies of coherent atom optics that exploit the ability to trace the evolution of nearly monochromatic matter waves for long times; studies of scalar Bose-Einstein condensates within novel trapping geometries; studies of spinor Bose-Einstein condensates and other quantum gas mixtures within large three-dimensional volumes and under homogeneous conditions; studies of strongly interacting atomic and molecular quantum gases over a wide range of densities; and applications of nonlinear optics for quantum communication and quantum information science, including storage of quantum states of light within a space-based platform. As the constellation of space-based ultracold atom experiments continues to advance and proliferate, and as the scientific community continues to identify creative uses of these experimental facilities, the scientific domain of such space-based ultracold-atom research is bound to expand further.

Even with these positive developments, one should ask whether this approach to ultracold atomic physics research in space will lead us to the greatest potential scientific impact. That is, the current approach involves taking terrestrial ultracold atomic physics experimental setups, packaging them up to be as robust and self-sufficient as possible, and then installing them aboard space-based platforms -- the International Space Station, or, in the future, a Lunar Gateway facility or even dedicated freely flying spacecraft. In this approach, we neglect the fact that space is different, and therefore the experimental techniques and setups for ultracold-atom physics in space should also be different. Experiments on Earth are performed within laboratories that have a limited size and are near atmospheric standard pressure and temperature (apparatuses within the laboratory, of course, may contain much lower pressures, and operate at lower temperatures). This fact "bakes in" specific choices and limitations of experimental techniques for producing, trapping, cooling, manipulating, and measuring ultracold atomic gases. Outer space is an entirely different laboratory setting, with essentially unlimited volume, extremely low background pressure, extreme departures from room temperature, and, of course, free fall. How should ultracold atomic physics be conducted within this different laboratory setting, and what novel directions for research does this different approach open up?

In this paper we explore some of the limits for ultracold gases and Bose-Einstein Condensates (BEC) that might be extended in space. These include the number of atoms in a condensate, the physical volume of a condensate, the lowest temperature achievable, the longest free propagation time and distance, and the largest and most separated quantum superposition of atoms.

Improving these limits will increase the sensitivity of virtually all proposed missions involving cold atoms, and hence we believe its vitally important to explore these issues as fully as possible before investing heavily in dedicated missions aimed at addressing specific scientific questions.

**2. Limits to Bose Condensation**

We begin by discussing limitations to experiments performed to date on ultracold atomic gases. For specificity, let us focus here on Bose-Einstein condensates with an eye toward their eventual application as coherent matter-wave sources for atom interferometry and other experiments. The largest BEC of alkali gases made to date have about $10^8$ atoms. [8] Terrestrial BEC densities are typically limited to the neighbourhood of $10^{14}$ atoms/cc due to losses arising from three-body collisions. These losses occur preferentially for the coldest atoms, resulting in an effective heating mechanism as well as a loss mechanism. High densities are essential for efficient evaporative cooling, but achieving high atom number typically requires relaxation of the trap strength in the final stages of cooling in order to reduce three-body losses. Since these losses scale as the square of density, it is possible to find a density range with reduced losses that still supports a high level of elastic two body collisions needed for thermalization. [9] On Earth, however, one is constrained by the requirement to support the condensate against gravity, limiting how weak the trap can be made. In microgravity, one can dramatically weaken the confining potential, allowing 3 body losses to be arbitrarily small. Weakening the trap will lower the thermalization rate however, so that achieving the highest number condensates will require both microgravity and improvements in vacuum compared to state-of-the-art condensate machines. [8,10]

Microgravity has already demonstrated its utility in achieving low effective temperatures. Here "effective temperature" is simply the temperature equivalent of the average kinetic energy of expansion, a useful metric particularly for precision measurements requiring very long interrogation times. Two microgravity experiments have achieved effective temperatures in the range of a few tens of picokelvins, and there do not appear to be fundamental limits that would prevent us from achieving sub-picokelvin temperatures. [3,11] These experiments employed a technique known as delta-kick cooling (DKC), [12] where atoms are allowed to freely expand before a harmonic potential is briefly applied to bring them to a halt. Ultimately, a limit to achievable temperature in a given apparatus will arise from the apparatus size, which limits expansion time.

*Ultracold atoms in the vacuum of space*

One of the most striking advantages of conducting ultracold atomic physics experiments in space is the access to the vacuum of space. Ultracold atomic gases remain ultracold within the laboratory only by being suspended within an ultrahigh vacuum (UHV), providing a vacuum thermos that isolates the micro- or nano-kelvin atoms from the infernally hotter temperature of the surrounding experimental apparatus. The quantum gas cannot exist for even a nanosecond ``outdoors,'' i.e. within the room-temperature and atmospheric-pressure laboratory in which the experiments are conducted. Pressures at the level of $10^{-11}$ torr, that is, around 13 orders of magnitude below atmospheric pressure, are achieved within vacuum chambers constructed with specifically chosen materials and prepared by specialized techniques to reduce outgassing and eliminate vacuum leaks. There is a constant struggle to maintain optical access with high numerical aperture and through high optical grade windows, and, for some applications, to provide space for the gas to propagate and expand, while also maintaining ultrahigh vacuum, providing high conductance to vacuum pumps, and so on.

The vacuum of outer space offers a dramatically different environment. If we step outside the shirtsleeve environment of the International Space Station, a Lunar Gateway, or other manned spacecraft, we encounter ``outdoor'' laboratory conditions that are much more compatible with ultracold atomic physics. The pressure outside the ISS -- low-Earth orbit (LEO) at an altitude of around 400 km -- is several times $10^{-8}$ torr. This matches the pressure of vapor cell 2D and 3D magneto-optical traps. At higher altitudes within LEO, say 600 km or higher, the pressure drops to the $10^{-10}$ torr range, near the UHV requirements of cold-atom experiments with vacuum lifetimes in the range of 10's of seconds [see below]. Beyond LEO, the background pressure is lower still, eventually reaching extreme high vacuum levels that are hard to achieve in any terrestrial laboratory, let alone within an ultracold-atom experiment. We should note that within an experimental UHV chamber, the residual gas is neutral, whereas in various ``outdoor'' environments in space, a major component of the residual gas is ionized. [13] The cross-section for charged-neutral collisions is larger, by perhaps two orders of magnitude, than neutral-neutral collisions owing the longer range of charge monopole-dipole interactions ($1/r^4$, with r being the particle separation) as compared to dipole-dipole interactions ($1/r^6$). [14] As such, the vacuum requirements in space environments may be more stringent than in terrestrial UHV environments in order to achieve the same ultracold-atom trapping lifetimes. The pressures stated above are from the NRLMSISE-00 model of the Earth's upper atmosphere, [15] and assume mean solar activity.

Taking ultracold atomic physics outdoors would open dramatic new scientific possibilities. An ultracold gas could be made to expand to enormous volumes, spanning orders of magnitude in density down from the $\sim 10^{14}$ cm$^{-3}$ upper limit set by three-body losses. This range offers novel settings for studying two- and three-body collisional resonances and preparing gases with extremely low temperature, kinetic energy, and interaction energy. [16] Coherent matter waves ("atom lasers"[17]) could be directed to propagate over enormous distances: imagine an atom-laser source that emits coherent matter wave beams through space, and a detector that sits many meters away that receives the laser pulse and probes the spatial coherence of the matter wave. Atom interferometers could operate with enormous baseline distances and areas, allowing for enhanced sensitivity to acceleration, rotation, gravitation including violations of the weak equivalence principle, and the recoil frequency. [7]

In the outdoors of space, ultracold atomic gases can be prepared far from any experimental apparatus or other objects. In terrestrial laboratories, and also within the ISS or other Earth-laboratory-scale experiments, ultracold atoms are prepared in the near vicinity of materials that produce stray gravitational fields, that radiate blackbody radiation that shifts atomic energy levels, or that produce other environmental disturbances. These disturbances influence the otherwise free propagation of atomic gases and matter waves. In contrast, the unlimited volume of outer space would allow one to manipulate atoms with instruments that are kept very far away, greatly reducing their stray influences on the quantum gas.

Further, it is tempting to consider quantum gases as highly sensitive probes of previously undetected particles and forces. As considered for example in the chameleon mechanism for dark energy [18], it is possible that measurement targets are obscured from a cold-atom quantum sensor that is ensconced within a vacuum chamber. Quantum gases in the outdoors of space, operating for example as sensitive hyperfine or optical frequency atomic clocks, magnetometers, force- or rotation-sensing interferometers, etc., might regain sensitivity to such targets.

A more prosaic application of cold atoms in space is simply as an excellent sensor of the vacuum itself: measuring the loss-rate of ultracold atoms from a magnetic trap depends only on the collision cross section, a fundamental atomic property. Hence cold atom vacuum sensors function as both an absolute sensor and primary vacuum standard [19,20].

These advantages beckon us to move ultracold atomic physics outdoors into space. In the following we discuss several options for possible cold atom experiments.

*Lunar surface*

The lunar surface might be the ideal place to begin experiments utilizing the vacuum of space. Here an open structure holding the required optics could be set up stably on the lunar surface with no need for an independent spacecraft. While the moon's gravity is much greater than what can be achieved in freefall, it is still weak enough to achieve long observation times and significantly weaker traps. An atomic fountain on the Moon would achieve free-fall observation times roughly 2.4 greater than on Earth. And, of course, in a system with no vacuum system, there is no fundamental limit as to how tall the fountain can be.

The lunar atmosphere is currently on average about $2 \times 10^{-12}$ torr during the lunar night, [21] which is significantly better than that obtained in CAL, though still not as good as state of the art terrestrial vacuum systems. This atmosphere may be impacted somewhat by a return of humans to the moon — the total mass of the lunar atmosphere is only about 25000 kg, while each of the Apollo lunar landers carried over 10000 kg of fuel (the measured impact of each Apollo mission on the lunar atmosphere was about 0.2 lunar atmosphere masses, which dissipated on photoionization timescales of less than a few weeks). [21] Outgassing from manmade equipment and human activities, such as construction or mining, could be additional sources of lunar atmosphere. Monitoring the highly variable lunar atmosphere with cold atoms would be of considerable scientific and technical interest and could be an important secondary goal of a lunar mission. [22,23] An ideal location for such a facility would be a shadowed location near the south pole, where we would expect even better vacuum, and in addition the system would not be exposed to the dramatic temperature swings found elsewhere on the lunar surface.

Lacking a geodynamo, the moon has a very small magnetic field which varies widely across the surface but is typically of order a few tens of nT, roughly 1000 times weaker than the Earth's field. [24] Hence, there would be no need for magnetic shielding or compensation for a typical ultracold atom experiment on the lunar surface. Again, cold atoms might prove an ideal way to study such weak fields, and potentially could be utilized to probe the nature of localized magnetic field anomalies on the moon, the origins of which are the subject of considerable interest. [25]

The effects of lunar dust, UV degradation of coatings and micrometeorite impacts on sensitive optics must be considered. As an example, the lunar reflectors deployed during the Apollo missions had their reflectivity degraded by over 90% after 44 years of operation (lunar dust is considered the most likely primary culprit). [26] These effects could be partially mitigated by the use of sleeves and shutters over optics, perhaps along with electrostatic fields to repel dust or mechanical cleaning. The harsh thermal conditions arising from the lunar day/night cycle will also need careful consideration, but are expected to be much less significant in permanently shadowed locations. Finally the lunar surface radiation environment must be taken into account. Because the biggest sources of radiation arise from the sun, [27] they are also expected to be reduced in shadowed locations.

*Wake shields in LEO*

A wake shield is essentially a metal plate that pushes residual gases out of the path of a rapidly moving spacecraft. As long as the velocity of the shield is substantially higher than the average velocity of thermospheric molecules, an ultra-high vacuum can be created in its wake. The Wake Shield Facility (WSF) was utilized on three Space Shuttle missions (STS-60, STS-69 and STS-80) and achieved vacuums of $10^{-10}$ torr behind a 3.7 m diameter stainless steel shield. Such a vacuum is barely adequate for a BEC experiment. However, this vacuum was likely limited by outgassing from the shield itself, and a properly vacuum-processed shield could achieve pressures as low as $10^{-14}$ torr, sufficient for most applications. [28]

For cold atom applications, we imagine a deployable optics assembly shielded behind a wake shield and tethered to a spacecraft or space station which would supply power and uplink/downlink capability. Active control of magnetic fields would likely be required to remove effects of the changing Earth field. DC electric fields, which can often be large in the vicinity of orbiting spacecraft, [13] would also require careful consideration, and possible mitigation. Such an instrument might be interesting as a basis for developing an atom interferometer for gravity gradient measurements of the Earth. [29]

As we have discussed above, one of the prime motivations for performing ultracold experiments in the vacuum of space is to allow for very large apparatuses to be deployed. In such cases the expenses of launching a traditional metal or glass vacuum chamber would be prohibitive. An alternative that might be considered in locations where the vacuum of space is not significant for a particular application might be a deployable thin-film mylar or polyimide bag supported by a light weight ribbing that would easily hold up against the outside pressure of $10^{-9}$ torr. A vacuum pump within that thin-membrane chamber, e.g. titanium sublimated against a portion of the inner surface, would readily bring pressure down to the UHV regime: the effective area of the pumping surface, which could be on the scale of square meters, need only be several hundred times larger than the cross-section of all the holes in the walls of the vacuum chamber. This thin vacuum chamber could be optically transparent and thin enough so as to neither attenuate nor distort light beams sent into and through the chamber.

*Interplanetary space*

Near ultimate vacuum can be found in interplanetary space. While the costs of a dedicated free-flying mission orbiting at, say, one of the Earth-Sun Lagrange points might be prohibitive, several of the most demanding applications of cold atoms in space may need to utilize this environment. These include missions that aim to search for gravity waves, and for dark energy or dark matter candidates. [30] In this environment, the lifetime of quantum matter can be hundreds of times longer than achievable on Earth.

*Passively cooled high temperature superconductors*

Scaling up a BEC apparatus in general requires a large increase in mass and power. Making use of the vacuum of space, as we've discussed above, helps the mass budget significantly by avoiding the mass of vacuum pumps and chambers, with the mass of most other components (cameras, computers, electronics and lasers, etc.) only increasing modestly, if at all.

On the power side, the main components that require more power as you scale the system are lasers and magnetic field current drivers. Unfortunately, the current I required to produce a given magnetic field curvature scales as $I/S^3$, where S is the characteristic size of the system. [31] Hence, doubling the size of a harmonic magnetic trap will generally require a 64-fold increase in power.

We can largely eliminate this power increase by converting to superconducting magnetic coils. Here we propose to utilize another feature of the space environment, namely the ability to radiatively couple to the roughly 5K background temperature of space (at 1 AU from the sun this is somewhat above the cosmic microwave background due to zodiacal dust). Passive cooling to below 30 K is readily achievable in space, [32,33] though care is needed to shield from Sun, Earth and Moon shine. Wires made from YBCO (92 K transition temperature) BSCCo (108 K) and REBCO (REBa$_2$Cu$_3$O$_x$, where RE = Y, Gd) have been utilized in high current magnet applications producing steady-state fields over 45 T, [34] much larger than needed for the current application.

The cold temperatures of outer space can also be used to shield sensitive experiments from blackbody radiation, and could also be used to passively cool laser systems and other components.

## 3. Why explore these limits?

One of the primary reasons for studying cold atoms in space is to support the development of exquisitely sensitive quantum sensors based on atom interferometry. Such sensors may be used to test fundamental theories of physics such as general relativity, search for exotic dark energy or dark matter candidates; or serve as observatories for gravitational waves. Atom interferometry can also be used to monitor the Earth's gravitational field, aid prospecting on Mars, or help navigate spacecraft.[7] For each of these applications, sensitivity is enhanced with larger atom numbers, colder temperatures and longer lifetimes. It is important that we fully explore and understand these limits before embarking on costly space missions focused on a specific scientific objective.

More speculatively, achieving atom numbers well above the current state-of-the-art might open new possibilities for studying fundamental physics. Standard theories of quantum mechanics and gravity are expected to break down as we probe nature at the Planck scale. [35] While the Planck time, energy, and length are well beyond the reach of foreseeable experiments, the Planck mass ( $m_p = \sqrt{\hbar c/G}$ or 2.2 x10$^{-8}$ kg) appears to be more accessible. The idea is that as we examine manifestly quantum objects that approach this mass, such as pure condensates, superpositions or entangled ensembles, we may see deviations from current theory that could point towards a quantum theory of gravity. Achieving the Planck mass would require a condensate of $1.5 \times 10^{17}$ rubidium atoms, and is unlikely to be obtained even with the enhancements discussed in this paper. However, a number of authors have suggested that precision measurements may be able to observe deviations at much smaller atom numbers: for example, Roger Penrose and colleagues have proposed studying a single BEC in a superposition of two locations which could test a quantum gravity proposal in which wavefunction collapse emerges from a unified theory as an objective process, resolving the longstanding measurement problem of quantum mechanics. [36] In this case, the authors considered a Cs BEC with $4 \times 10^9$ atoms. Others have proposed looking for non-Gaussianity as a signature of quantum gravity in a sample of $10^9$ atoms left to self-interact gravitationally for several seconds. [37] Both experiments would likely require long duration microgravity, even if a terrestrial means of achieving the desired atom numbers and temperatures was found.

## 4. Technology roadmap

A significant technology development effort will be required to achieve the goals we've discussed in this paper. We outline in this section a few of the most important areas in which we believe resources should be deployed.

Efficient, high power laser systems: Collecting large atom numbers requires large beam diameters and high-power lasers. We will probably not be able to surpass limits achieved on Earth [8,10] without developing multi-Watt, space-qualified laser systems suitable for laser cooling at near IR wavelengths. Passive cooling and intermittent operation may help to reduce power requirements.

Deployable optical systems: For experiments utilizing the vacuum of space, we envision a folding structure for mounting optics and magnetic coils. Rather than designing a system that can maintain stringent alignments in the external space environment, we would aim to develop active steering of optics.

Compact very high flux cold atom sources: In the short term, 2D MOT-derived cold atom sources appear to be the best option when one tries to optimize both flux and size, weight and power (SWaP). Zeeman slowers incorporating superconducting magnets may be a more long-term solution.

Space based passively cooled high temperature superconducting (HTS) magnets: While HTSs have been tested in a space environment [37], passively cooled HTSs have not yet been demonstrated in space. A ground program should focus on the choice of HTS materials, designs of magnets for cold atom applications, and the thermal design needed for passive cooling in different space environments.

## 5. Costs

A program to extend the limits of ultracold atoms in space will require a significant (on the order of $1-2 million/year) ground research program for the coming decade, in conjunction with a series of space missions. An initial focus should be on

achieving the current state of the art for atom numbers and lifetimes of Earth-based experiments in a space-flyable instrument. We expect a next-generation facility, such as the Quantum Explorer [38], to largely accomplish this goal. A follow-on mission deployed on the moon might be an ideal way to develop technologies for an instrument that would utilize the vacuum of space.

The Lunar Ultracold Neutral Atom Research (LUNAR) facility would consist of an open support structure to hold optics, atomic sources, and magnetic coils, along with associated electronics and lasers. The complexity of the instrument would be nearly identical to the existing Cold Atom Lab (CAL), currently operating onboard the ISS, though it would at least initially only incorporate a single atomic species. We expect significant additional design effort is necessary for the thermal, optomechanical and possibly electronic systems. In addition, we would likely reduce the operation time from the 3+ years demonstrated by CAL to something more like a 120-day mission duration. Beyond being an important technology demonstration, such a mission would give new insights into the Moon's dynamic atmosphere, and would also allow for a short, highly focused, research program incorporating very large condensates.

Assumptions obviously include the existence of a manned facility on the moon that would provide power and data with standard interfaces (similar to the ISS), and that costs of transporting the instrument to the surface of the moon, along with crew time are not included in the estimate.

## Acknowledgements

The research was carried out at the Jet Propulsion Laboratory, California Institute of Technology, under a contract with the National Aeronautics and Space Administration (80NM0018D0004). © 2022. California Institute of Technology. Government sponsorship acknowledged.